\documentclass[a4paper,11pt]{article}
\usepackage{geometry}
\usepackage{amsmath}
\usepackage{amssymb}
\usepackage{sidecap, caption}
\usepackage{graphicx}
\usepackage{latexsym}
\usepackage{times} 
\usepackage[english]{babel}
\usepackage{fancyhdr}
\usepackage{wrapfig}
\usepackage{subcaption}
\usepackage{fontawesome5}
\usepackage{multirow}

\usepackage{booktabs}
\usepackage{multirow}
\usepackage[table,xcdraw]{xcolor}

\definecolor{lightblue}{HTML}{ADD8E6}
\definecolor{darkblue}{HTML}{00008B}
\definecolor{lightgreen}{HTML}{90EE90}
\definecolor{darkgreen}{HTML}{006400}

\sidecaptionvpos{figure}{c}

\usepackage[colorlinks=true,linkcolor=blue]{hyperref}

\hypersetup{
  colorlinks,
  citecolor=ULBblue,
  linkcolor=ULBblue,
  urlcolor=ULBblue}

\definecolor{ULBblue}{rgb}{0,0.2196,0.5765}

\addto{\captionsenglish}{\renewcommand{\abstractname}{Executive Summary}}

\fancyfootoffset{1cm}
\fancypagestyle{plain}{%
\fancyhead{}
\cfoot{}
\fancyfoot{}
\rfoot{\thepage}
\lfoot{{\small EVS35 International Electric Vehicle Symposium and Exhibition}}

}

\renewenvironment{abstract}
 {
  {\bfseries \large{\abstractname}}
  \par
  \vspace{10pt}
  \normalsize
 }

\addto\extrasenglish{%
}
\addto\extrasenglish{%
}
\addto\extrasenglish{%
}

\title{\Large \emph{35th  International Electric Vehicle Symposium and Exhibition (EVS35)}\\ \emph{Oslo, Norway, June 11-15, 2022}\\ \hspace{10pt}\\ \LARGE\bf 
Estimation of Public Charging Demand using
Cellphone Data and Points of Interest-based
Segmentation} 

\author{
{\large Victor Radermecker$^1$, Prof. Lieselot Vanhaverbeke}\\
{\small $^1$\em victor.timothee.radermecker@vub.be, MOBI Research Group,} \\ 
{\small \em Vrije Universiteit Brussel,  Pleinlaan 2, 1050 Brussels, Belgium} \\
}

\date{}
\begin{document}

\setlength{\parindent}{0mm}
\baselineskip 10pt

\maketitle

\rule{\textwidth}{1pt}
\begin{abstract}

The race for road electrification has started, and convincing drivers to switch from fuel-powered vehicles to electric vehicles requires robust Electric Vehicle (EV) charging infrastructure. This article proposes an innovative EV charging demand estimation and segmentation method. First, we estimate the charging demand at a neighborhood granularity using cellular signaling data. Second, we propose a segmentation model to partition the total charging needs among different charging technology: normal, semi-rapid, and fast charging. The segmentation model, an approach based on the city’s points of interest, is a state-of-the-art method that derives useful trends applicable to city planning. A case study for the city of Brussels is proposed. \\

\end{abstract}

\rule{\textwidth}{1pt}
\vspace{10pt}

\section{Introduction}
Climate change is probably the premier hurdle ever faced by our societies, and as engineers and scientists, it is our responsibility to drive forward research towards a sustainable world. Most cities worldwide have set high environmental ambitions for decarbonizing their transport network. 
As public charging stations occupy the already scarce public space, governments need a comprehensive understanding of their cities’ charging demand patterns to ensure the optimal location of these newly built charging points.  \\

Today, three major EV charging technologies exist: normal, semi-rapid, and fast. They differ in the type of electric current (AC/DC) and the maximum power delivered by the terminal, which results in different charging speeds for drivers. While the fastest stations can meet the needs of more drivers and reduce range anxiety, they are the most expensive to build and can cause electric grid instabilities. \\

Many scientific studies have been carried out to optimally locate these charging stations in urban areas. However, most researchers have not focused on the segmentation of the charging demand between these different technologies. This is problematic as city planning officials need to know where which types of stations would be most suitable. This article proposes a method to estimate and segment the demand into the aforementioned three charging speeds.

\section{Literature Review}
The literature offers many articles on optimal locations of public EV charging stations. First, \autoref{independent} presents different studies which do not distinguish the different charging technologies. Most of these articles combine demand estimation methods with optimization models to derive the optimal locations of charging points. Second,  \autoref{dependent} summarizes some studies addressing roll-out strategies for specific charging speeds. Although some papers are starting to emerge, very few authors have investigated these issues in detail.  

\subsection{Charging Type Independent Approaches}\label{independent}
\subsubsection{Demand Estimation}
Given its importance in optimization algorithms, EV charging demand estimation has been examined by researchers worldwide. Many authors took advantage of Cellular Signaling Data (CSD) to estimate charging demand. Among them, Jia et al. \cite{B1} proposed an approach to locate public charging stations using reconstructed EVs trajectories derived from simulation. The simulation is done through a mesoscopic simulation tool that can reconstruct a trajectory based on multiple checkpoints \cite{NEXTA}. Although presenting promising results, this study requires CSD at an individual level, which is difficult to collect. For instance in Europe, such data mining goes against the GDPR (General Data Protection Regulations). Whereas Jia et al. \cite{B1} assumed a constant charging probability for EV drivers, Cao et al. \cite{B2} proposed a probabilistic model to simulate EV drivers’ charging habits. Their model included various factors, including the State of Charge (SOC), the next driving mileage, the travel time value, and economic factors. \\  

Socio-demographic indicators are popular variables to model the charging demand. He et al. \cite{B8} proposed to use institutional and spatial factors to model Beijing's charging demand. They have identified six key socio-demographic attributes: income, vehicle ownership, educational level, age, gender, and family size. Frade et al. \cite{B17} used a similar approach for Lisbon, but they distinguished nighttime and daytime charging. Nighttime demand was derived using the number of cars per household, whereas daytime demand was estimated from the volume of employment.  \\

Travel surveys and traffic simulations are also interesting methodologies. Efthymiou et al. \cite{B4} used traffic simulation models to build a travel Origin-Destination (OD) matrix employing data collected by phone surveys. Similarly, Baouche et al. \cite{B6} took advantage of household travel surveys to build an OD matrix for the city’s neighborhoods. They used a routing API to compute the driving distance under realistic traffic conditions and build an energy demand OD matrix through an EV consumption model \cite{B18} \cite{B19}. Cavadas et al. \cite{Z2} exploited a mobility survey that questioned drivers about their daily trips on business days. They built an energy OD Matrix by converting these trips into EV consumption under specific assumptions.  \\

Finally, some authors assumed that car-parking demand is proportional to EVs' charging demand in urban areas. Chen et al. \cite{D3} aggregated a dataset of around 50 000 person trips and combined it to land use characteristics. Using this input data, they trained an Ordinary Least Squares (OLS) regression to ultimately predict the total parking times and derive the charging demand. 

\subsubsection{Optimization Models}
Once the charging demand is estimated, optimization models allow locating optimal sites for stations. Researchers such as Zuo-Jun et al. \cite{B3} have extensively aggregated the existing optimization models. To name a few, Upchurch et al. \cite{Z5} introduced a capacitated refueling location model (CRLM) to locate EV charging stations. The original uncapacitated model assumed that the presence of a refueling station is sufficient to supply all vehicles intersecting the station node, which was not realistic for EVs. Cavadas et al. \cite{Z2} proposed a mixed integer programming (MIP) model for locating charging stations. The model considered not only the drivers' parking locations but also their daily activities. \\ 

Efthymiou et al. \cite{B4} proposed a genetic algorithm to estimate the optimal location of charging stations. The advantage is that their algorithm provides valuable results with limited data requirements. Sweda and Klabjan \cite{B5} have created an agent-based decision support system. Its end goal was to identify patterns in residential EV ownership and enable strategic deployment of new charging infrastructure.  Xi and Sioshansi \cite{Z6} have developed a simulation–optimization model which determines optimal EV charging stations' location by maximizing their popularity among EV drivers.

\subsection{Charging Type Dependent Approaches}\label{dependent}
Only a few studies focused on EV charging infrastructure for specific charging technologies. Zhang and Iman \cite{D2} computed suitability scores for normal charging stations in urban areas. Their approach relied on using Geographic Information System (GIS) data to determine the sites' suitability for hosting normal EV charging stations. By aggregating various input factors such as land use, demographics, and employment centers, they derived suitability scores for all stations in Utah (US). Agent-based models have also been explored to investigate charging stations roll-out strategies for semi-rapid and fast technologies. Hoed et al. \cite{Z8} simulated a population of diverse agents (citizens, foreigners, taxis, and shared vehicles) to model the multiplicity of charging behaviors within cities. Regarding fast charging, Morrissey et al. \cite{D1} investigated a widely accepted roll-out strategy, which consists of creating electric corridors on major interurban motorways to connect cities. They assessed the adequacy of this roll-out strategy, which in theory, should allow longer trips to be undertaken and increase EV users' potential journey ranges.  \\

Some authors explored optimization methods for specific charging technologies. For instance, Wang and Chuan-Chih \cite{Z7} have expanded the concepts of set-coverage and maximum-coverage to express a capacitated location model that deals with multiple types of charging stations using a MIP algorithm. The main contributions of this study are that it relaxes the constraints of using a single kind of recharging station. The results highlight that a combination of charging technologies, although it requires a higher budget, leads to greater convenience for travelers. Others have also explored optimization models to locate charging stations for specific fleets, such as taxis, shared cars, and buses \cite{Z3} \cite{Z4}.

\subsection{A new approach}\label{new}
Numerous studies focused on estimating the total EV charging demand to apply location models. However, these approaches do not sufficiently account for the various complementary charging technologies. This article proposes an innovative methodology, which, by segmenting the charging demand into the different charging speeds, allows to re-use the technology-independent models presented in \autoref{independent} for specific charging speeds, and hence helps provide actionable insights to city planning authorities. Our method is a two-step approach. First, we propose a demand estimation method, which relies on aggregated CSD, and thus respecting GDPR. Second, we take advantage of GIS data to assess each neighborhood's primary function and compute its needs for each charging type.  \\

The paper is structured as follows. Next, we present the methodology, which outlines our two-step method to estimate and segment public charging demand. To illustrate the model, we apply it to the case study of Brussels, Belgium. We present the datasets used and the various models. Results are summarized and discussed to highlight how they can provide actionable insights to planning authorities. Finally, we present conclusions of the work, identify its limitations and highlight future areas for continued investigation. \\



\section{Methodology}
As explained above, our method consists of two main steps. First, \autoref{demand_estimation} details the EV public charging demand estimation. Second, \autoref{demand_segmentation} develops the segmentation model.

\subsection{Demand Estimation}\label{demand_estimation}
\subsubsection{Methodology Overview}
\begin{figure}[h!]
    \small
    \centerline{\includegraphics[width=0.75\linewidth]{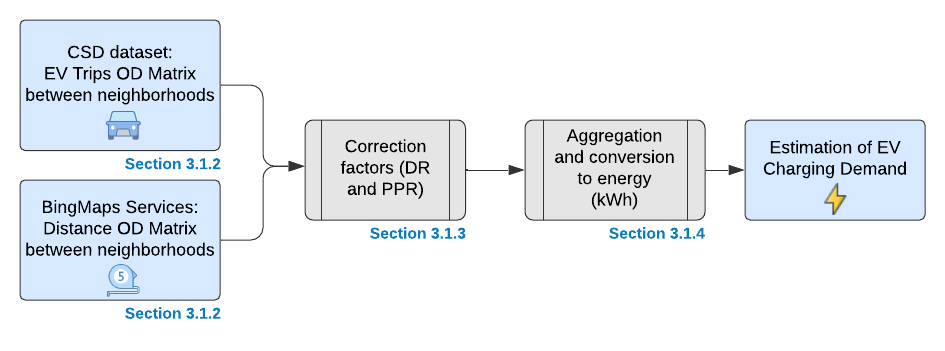}}
    \caption{Process of EV charging demand estimation for Brussels. \autoref{od_matrix} estimates the number of trips between districts based on CSD. Then, it builds an OD matrix by computing the quickest driving distance between neighborhoods. \autoref{correction_factors} applies two correction factors, the DR and PPR. \autoref{conversion} converts OD matrices into neighborhood-level energy demand in kWh.}
    \label{fig:demand-estimation-process}
\end{figure}

The goal of \autoref{demand_estimation} is to estimate the EV charging demand at neighborhood granularity for Brussels. The city is divided into 145 neighborhoods 
with each one having an average area of 0.51$km^2$. The whole process is presented in \autoref{fig:demand-estimation-process}. First, we take advantage of aggregated CSD to build an OD matrix revealing the number of trips between Brussels neighborhoods. In \autoref{od_matrix}, we convert these number of trips into driving distances using BingMaps Services. \\

Second, two correction factors are applied on top of previous results to make the model more realistic. (1) An analysis of mobility in Brussels leads the way towards the first improvement, the Driving Ratio (DR). This step removes walking, bike, and public transport trips from the data by estimating the ratio of car trips to other transportation means per neighborhood. (2) A socio-demographic study of private parking opens to the second refinement, the Private Parking Ratio (PPR). By estimating the number of citizens owning a private parking lot, we can infer the share of the population having access to a private at-home charger, thus not relying on public charging infrastructure. These factors are presented in \autoref{correction_factors}. \\

Finally, the total demand is aggregated at a neighborhood level and converted into energy demand in kWh. These conversions are based on the EV energy consumption averages established by the International Energy Agency (IEA). 

\subsubsection{Origin-Destination Matrix}\label{od_matrix}

\begin{wrapfigure}[16]{r}{0.4\textwidth}
\small
    \includegraphics[width=1.0\linewidth,trim=0pt 0pt 0pt 45pt]{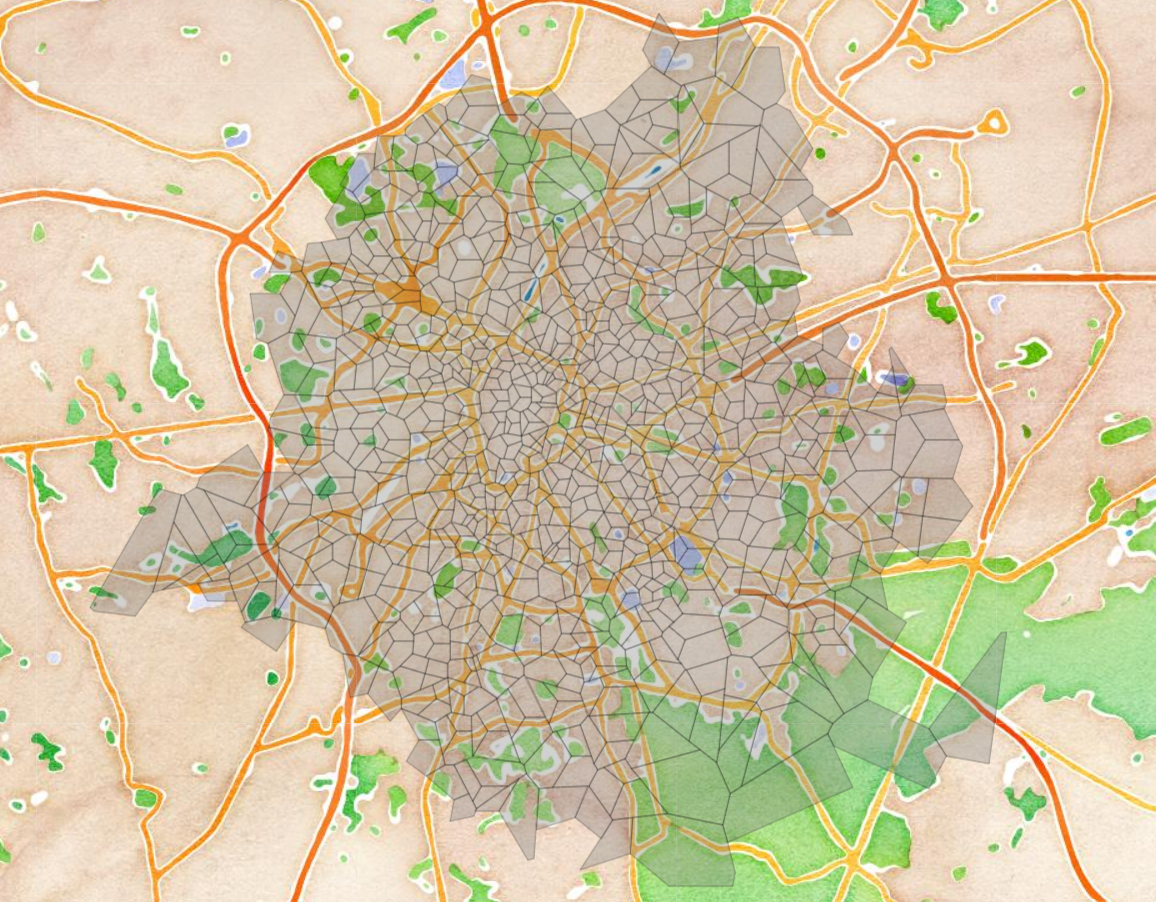}
    \caption{The 880 Proximus TACS covering the city of Brussels.}
    \label{fig:proximus-tacs}
\end{wrapfigure}

\paragraph{Dataset}
This dataset was provided by Proximus S.A., one of Belgium's largest telecommunication companies. It gives the total number of trips between specific origins and destinations pairs, defined as 'staying points' in a Technology Agnostic Cell Sector (TACS) network. Each TACS represents the unified coverage area of an antenna. A 'staying point' is defined as being in the same TACS for at least 30 minutes confirmed or 60 minutes estimated. Confirmed time is when the duration between the first and last network signals in the same sector is at least half an hour. Estimated time is computed as the time passed between two network signals incoming from different sectors. 

\vspace{0.5cm}

The data was collected on workdays, defined as all Mondays, Tuesdays, and Thursdays between 20/01/2020 and 20/02/2020. The final numbers are the average profile of these fifteen data collection days. The trips are divided into two categories: regular and irregular trips. A trip is defined as regular if it occurs not less than twice a week throughout at least four out the five weeks regardless of the time and day (including non-work days). Therefore, multiple daily trips are considered regular. All other trips are defined as irregular. As Proximus customers are not equally spread geographically throughout Brussels and Proximus is not the unique telecommunication services provider, an extrapolation factor is necessary to accurately reflect the population and prevent bias. The original dataset was aggregated at a neighborhood granularity by assigning each TACS to a specific neighborhood. The aggregation process uses the intersection over union indicator to associate the most probable district to each network signal. 

\paragraph{Distance OD matrix} As Brussels counts 145 neighborhoods, we have two matrices $M^r$ and $M^{ir}$ of dimensions 145x145, where each element $M_{ij}$ corresponds to the number of regular and irregular trips from neighborhood $i$ to $j$. We can convert these numbers of trips into a distance OD matrix by computing the average distance between all possible OD pairs. As Baouche et al. \cite{B6}, we will take advantage of an online geodata Application Programming Interface (API) to perform these computations. The BingMaps Services API provides an interface to perform tasks such as creating static maps with markers, geocoding addresses, retrieving imagery metadata, or creating routes.  \\

\begin{table}[h!] \small
    \centering
    \captionsetup{width=.84\textwidth}
    \caption{Distances (in $km$) between each pairs of the five neighborhoods presented above using the BingMaps API. Traffic conditions: Monday morning (7 AM).}
    \begin{tabular}{l||l|l|l|l|l|l} \small
     & Altitude 100 
     & Boondael
     & Vivier d'oie
     & Université
     & Observatoire
     & ...\\
    \hline
    \hline
    Altitude 100 & 0 & 5.937 & 5.906 & 5.430 & 3.386 & ...\\
    \hline
    Boondael & 5.616 & 0 & 5.930 & 1.486 & 4.590 & ...\\
    \hline
    Vivier d'oie & 6.181 & 4.868 & 0 & 6.478 & 3.594 & ...\\
    \hline
    Université & 5.516 & 1.582 & 7.609 & 0 & 4.809 & ...\\
    \hline
    Observatoire & 3.579 & 4.334 & 4.028 & 4.422 & 0 & ...\\    
    \hline
    ... & ... & ... & ... & ... & ... & ...\\
    \end{tabular}
    \label{tab:routes_length}
\end{table}

\autoref{tab:routes_length} shows the fastest driving distances computed using the BingMaps API for a subset of five neighborhoods in Brussels. We see that the diagonal elements of the OD matrix are null. Indeed, BingMaps returns a null distance for two identical points, in our case, the centroid's coordinates of the same neighborhood. However, as the CSD dataset includes non-null values for intra-neighborhood trips, we need to compute the average intra-neighborhood distance to fill the diagonal of $M_{ij}$. \\


The most accurate method is to numerically simulate the average distance for each specific neighborhood. We randomly sample pairs of points within each district and compute the relative euclidean distances. This process converges to the desired metric through the mean of all computed distances with sufficient iterations. These are straight-line distances, but people need to follow roads to reach their destinations in reality.
Boscoe et al. \cite{NationWideDistanceComparison} have analyzed the difference between straight-line and driving distances for more than 66 000 trips in 50 states of the United States. They highlighted highly correlated measures with a coefficient of determination
 ($ r^2 > 0.91$) and a detour index of 1.417. This index can be used to convert the distances as the crow flies to driving distances. 


    

\subsubsection{Correction Factors}\label{correction_factors}

\paragraph{Driving Ratio}

This first correction factor aims at answering the following question: what proportion of trips collected through the CSD accounts for trips made by cars? In other words, what is the balance of walk, bike, and public transportation trips? \\

\begin{wraptable}[12]{r}{9cm}     \small
\vspace{-0.7cm}
    \centering
    \caption{Proportions of each type of transportation means for 1 to 50km trips in Brussels \cite{plan-pieton} \cite{monitor}.}
    \begin{tabular}{l||l|l|l|l|l} \small
     & Drive
     & Public Tr.
     & Bike 
     & Walk 
     & Other \\
    \hline
    \hline 
    0 - 1 km & 17\% & 2\% & 14\% & 62\% & 5\% \\
    \hline 
    1 - 2 km & 40\% & 5\% & 26\% & 29\% & 0\%\\
    \hline 
    2 - 5 km & 59\% & 9\% & 19\% & 12\% & 1\% \\
    \hline
    5 - 10 km & 72\% & 11\% & 11\% & 4\% & 2\% \\
    \hline
    10 - 20 km & 78\% & 13\% & 5\% & 2\% & 3\% \\
    \hline
    20 - 50 km & 74\% & 22\% & $<$1\% & $<$1\% & $<$4\% \\
    \hline
    \end{tabular}
    \label{tab:pedestrian_habits}
\end{wraptable}
Mobility surveys, such as the recent \textit{Monitor} study for Brussels, are helpful to answer such questions. \autoref{tab:pedestrian_habits} presents the proportion of different transportation means for different trip lengths in Brussels (0 to 50 $km$) \cite{plan-pieton} \cite{monitor}. As we have estimated the distance of each specific OD trip, we need to assign a driving proportion to any length from 0 to 50 $km$. By multiplying the number of trips by this driving proportion, we get the corrected number of trips. This correction factor is called the DR. 

\vspace{0.5cm}

\begin{figure}[h!] \small
    \begin{minipage}[b]{0.485\textwidth}
        \centering
        \centerline{\includegraphics[width=\linewidth]{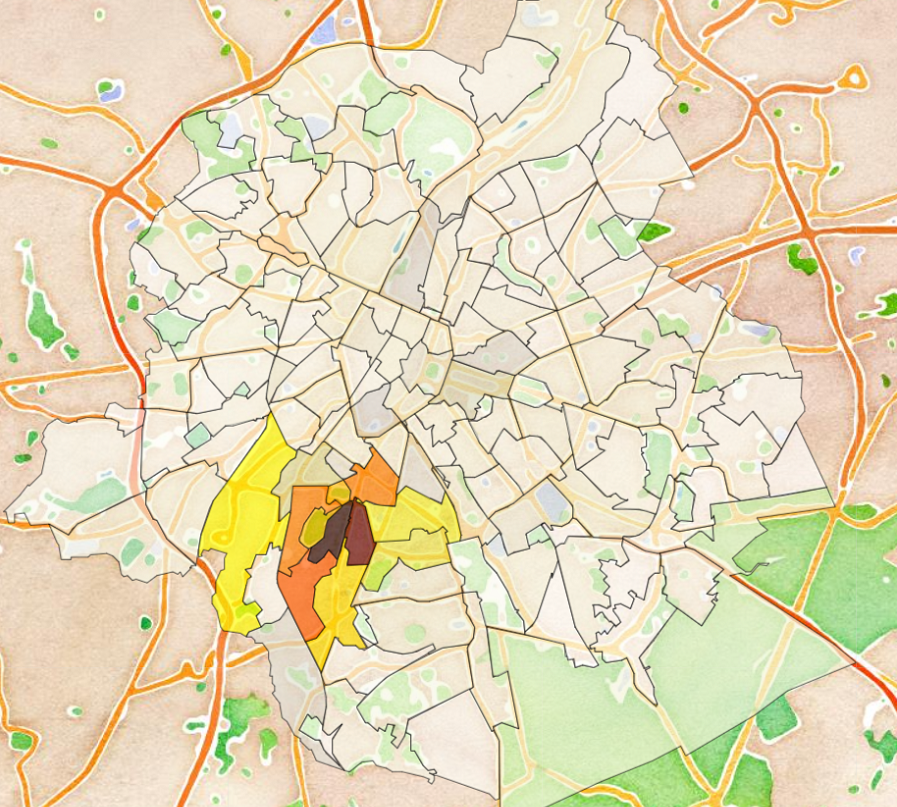}}
        \centerline{\begin{minipage}[b]{\textwidth}
        \centerline{\includegraphics[width=\linewidth]{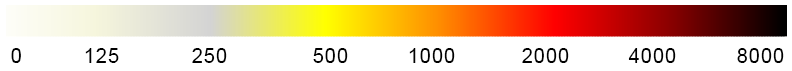}}
        \end{minipage}}
        \subcaption{Total number of trips (regular and irregular) originating from \textit{Altitude 100} before applying the DR correction factor. 
        \href{https://rawcdn.githack.com/victor-radermecker/BrusselsEVS35/ac9028504ddd608b4fd0d49400359416940f45de/img/regularALTITUDE100_Comparison1.html}{\textsc{Interactive Figure \faGithub}}}
        \label{fig:Altitude100-comparison-a}
    \end{minipage}
    \hfill
    \begin{minipage}[b]{0.485\textwidth}
        \centering
        \centerline{\includegraphics[width=\linewidth]{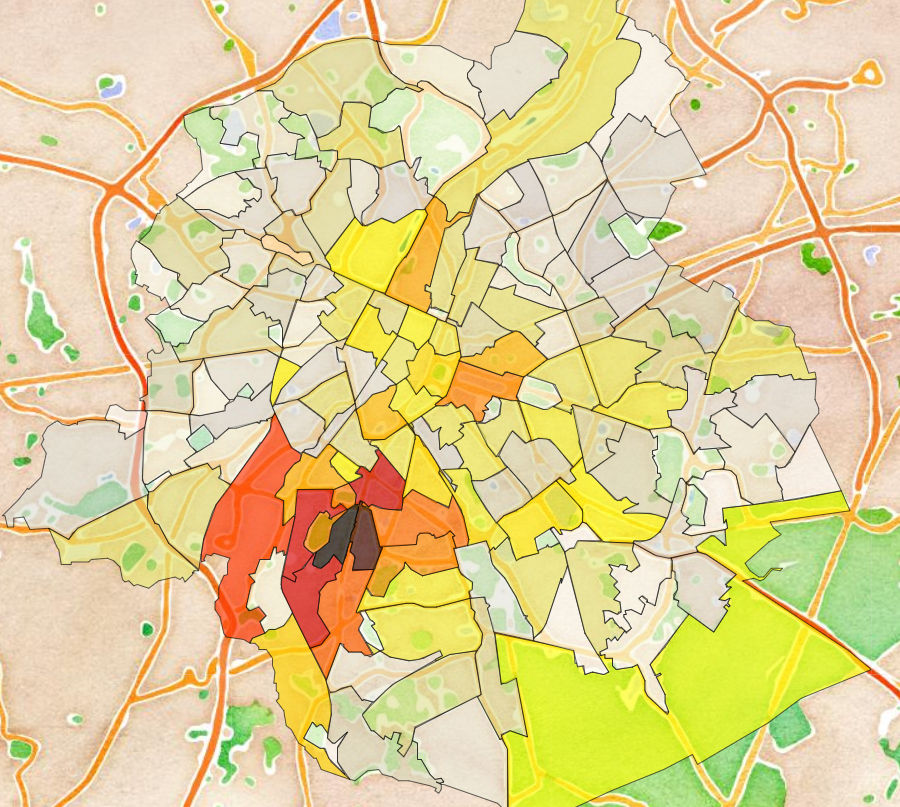}}
        \centerline{\begin{minipage}[b]{\textwidth}
        \centerline{\includegraphics[width=\linewidth]{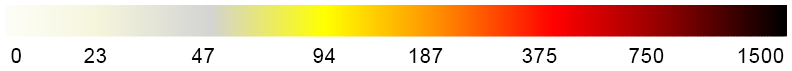}}
        \end{minipage}}
        \subcaption{Total number of trips (regular and irregular) originating from \textit{Altitude 100} after applying the DR correction factor. 
        \href{https://rawcdn.githack.com/victor-radermecker/BrusselsEVS35/ac9028504ddd608b4fd0d49400359416940f45de/img/regularALTITUDE100_Comparison2_Corr.html}{\textsc{Interactive Figure \faGithub}}}
        \label{fig:Altitude100-comparison-b}
    \end{minipage}
    \caption{Total number of trips, before and after applying the Driving Ratio correction factor.}
    \label{fig:Altitude100-comparison}
\end{figure}
Using the data given in \autoref{tab:pedestrian_habits}, we fit an exponential function, $f(x) = A(1-e^{-Bx})$, to estimate the driving ratio on a continuous spectrum. We use the non-linear least-squares optimization to find the best parameters $A$ and $B$ for this candidate function. The final sum of squares is $\sum_i (y_i - \hat{y_i}) \approx  0.001$, which is relatively low. The two estimated parameters are: $A = 0.759$, $B=0.466$. Results are shown on \autoref{fig:Altitude100-comparison}. 

\paragraph{Private Parking Rate}

The drivers' charging behavior is still difficult to anticipate, but among the various hard-to-predict factors, the PPR is a critical variable that must be reliably quantified. It is defined as the proportion of drivers who will own an at-home charging point and won't rely on public charging infrastructure. \\

Many recent studies concluded that most drivers prefered at-home charging. A recent investigation from Figenbaum showed that, in Norway, in 2017, 97\% of drivers living in semi-detached and detached housing were taking advantage of private charging points while only 3\% charged their car outside their home \cite{A12}. In the US, it is currently estimated that between 84 - 90\% of US EV drivers charge their car at home, a number that recently jumped with the COVID-19 pandemic \cite{Marketing2021Dec3}. Other studies in Australia also confirmed that home-charging, when possible, is the preferred solution for EV drivers \cite{SPEIDEL201497} \cite{C6}. Adding all governmental incentives for building private charging stations, we can confidently assume that all drivers having the possibility to install at-home charging points will do so. \\

\begin{wrapfigure}[7]{r}{.25\textwidth}
\begin{align}
  PPR_{i} &= \frac{\sum_j \sigma_j}{\tau_i \chi_i}
\end{align}
\end{wrapfigure}
The PPR is defined as the density of Private Access Roads (PARs) per household and neighborhood. PARs are defined as any private driveways, including garage doors and setback areas, private parking lots, etc. Large-scale inventories are regularly carried out by surveyors in Brussels, which include counting on-street parking spaces and photographing them in compliance with the GDPR. In Equation (1), $\sum_j \sigma_j$ represents the sum of all PARs in neighborhood $i$, $\tau_i$ is the population density and $\chi_i$ is the average household size of neighborhood $i$.

\subsubsection{Conversion}\label{conversion}

As the goal is to estimate the energy demand of each neighborhood, we need to convert these corrected distance OD matrices into energy values. The most accurate method to do this conversion is to use a consumption model simulator, as done by Baouche et al. \cite{B6}. However, we considered Brussels as a relatively flat city, which does not crucially need an altitude-based analysis for estimating EV consumption. Therefore, multiplying the average electricity consumption of an EV by the number of trips estimates the energy supply OD matrix. The International Energy Agency (IEA) provides reliable EV energy consumption metrics \cite{B15}. A passenger car consumes on average (on a road) between 0.20-0.24 kWh/km. In the analysis, we used the value of 0.22 $kWh/km$. 

\subsection{Demand Segmentation}\label{demand_segmentation}

\subsubsection{Methodology Overview}

In \autoref{demand_estimation}, we have estimated the public charging demand of EVs at a neighborhood level. Now, the goal is to segment this demand into normal, semi-rapid, and fast charging. \\

\begin{figure}[h!]
    \small 
    \centerline{\includegraphics[width=\linewidth]{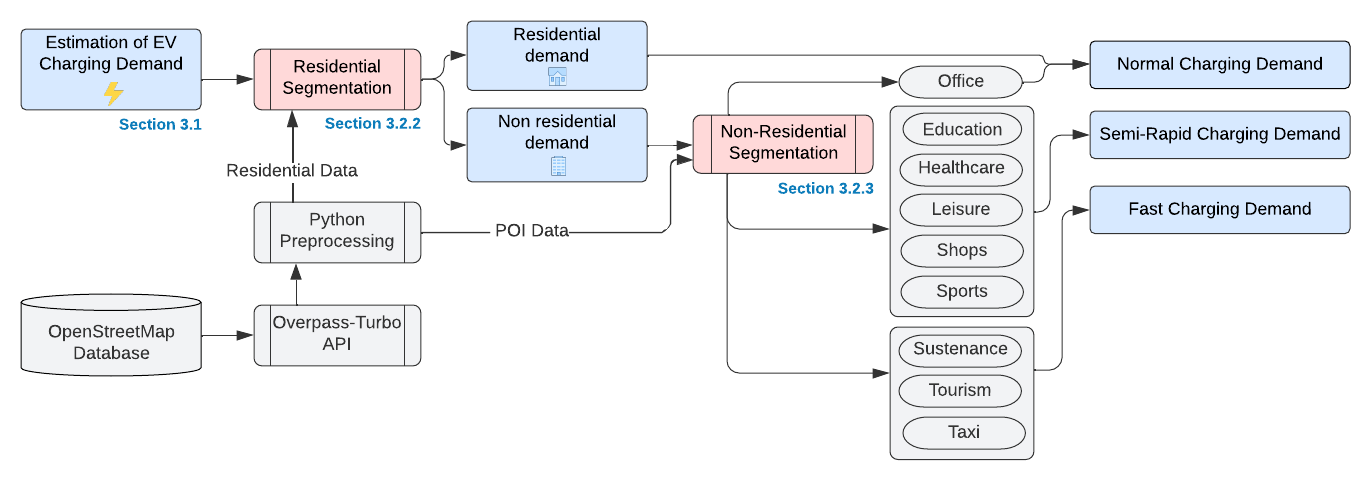}}
    \caption{\small This diagram shows the complete segmentation process. The two segmentation steps are highlighted in red. First, the residential segmentation called Highways Analysis is presented in \autoref{residential_segmentation}. Second, the non-residential segmentation called the POI-based segmentation is presented in \autoref{nonresidential_segmentation}. The charging demand at various stages is represented as blue cells. The data used in the model is extracted from the OpenStreetMap databases through the Overpass-Turbo API.}
    \label{fig:SegmentationFlowChartDetailed}
\end{figure}

The segmentation process, shown on \autoref{fig:SegmentationFlowChartDetailed}, consists of two independent steps. (1) We assume that EV drivers will always prefer normal charging in residential neighborhoods, which is cheaper than semi-rapid or fast charging. Therefore, the first step is to compute the proportion of residential charging demand for each neighborhood. This initial segmentation is called the Highway Analysis and is described in \autoref{residential_segmentation}. (2) The remaining non-residential charging demand is segmented in normal, semi-rapid, and fast charging using a state-of-the-art approach detailed in \autoref{nonresidential_segmentation}. As shown on \autoref{fig:SegmentationFlowChartDetailed}, the non-residential segmentation splits the demand according to the city's points of interest (POIs).

\paragraph{Data Mining}
This model requires a massive amount of up-to-date geodata to run effectively, making the data mining process critical. Different web maps services exist online such as GoogleMaps, BingMaps, and OpenStreetMap (OSM). The latter is a community-powered map that supplies many websites and applications with its data. It is entirely free to use yet maintains a remarkable level of accuracy thanks to the work of numerous volunteers and engineers who populate it \cite{AnnaDziuba2021}. Furthermore, OSM comes with Overpass Turbo, a web-based data filtering tool that makes the mining process more accessible. It is built on top of the Overpass API, a read-only API that extracts OSM map data and acts like an online shared database: the client sends a query to the API, which retrieves the corresponding dataset and exports it as a \textit{json} file. 

\newpage
\subsubsection{Residential Segmentation}\label{residential_segmentation}

\begin{wrapfigure}[20]{r}{0.5\textwidth} 
    \small
    \includegraphics[width=1.0\linewidth,trim=0pt 0pt 0pt 45pt]{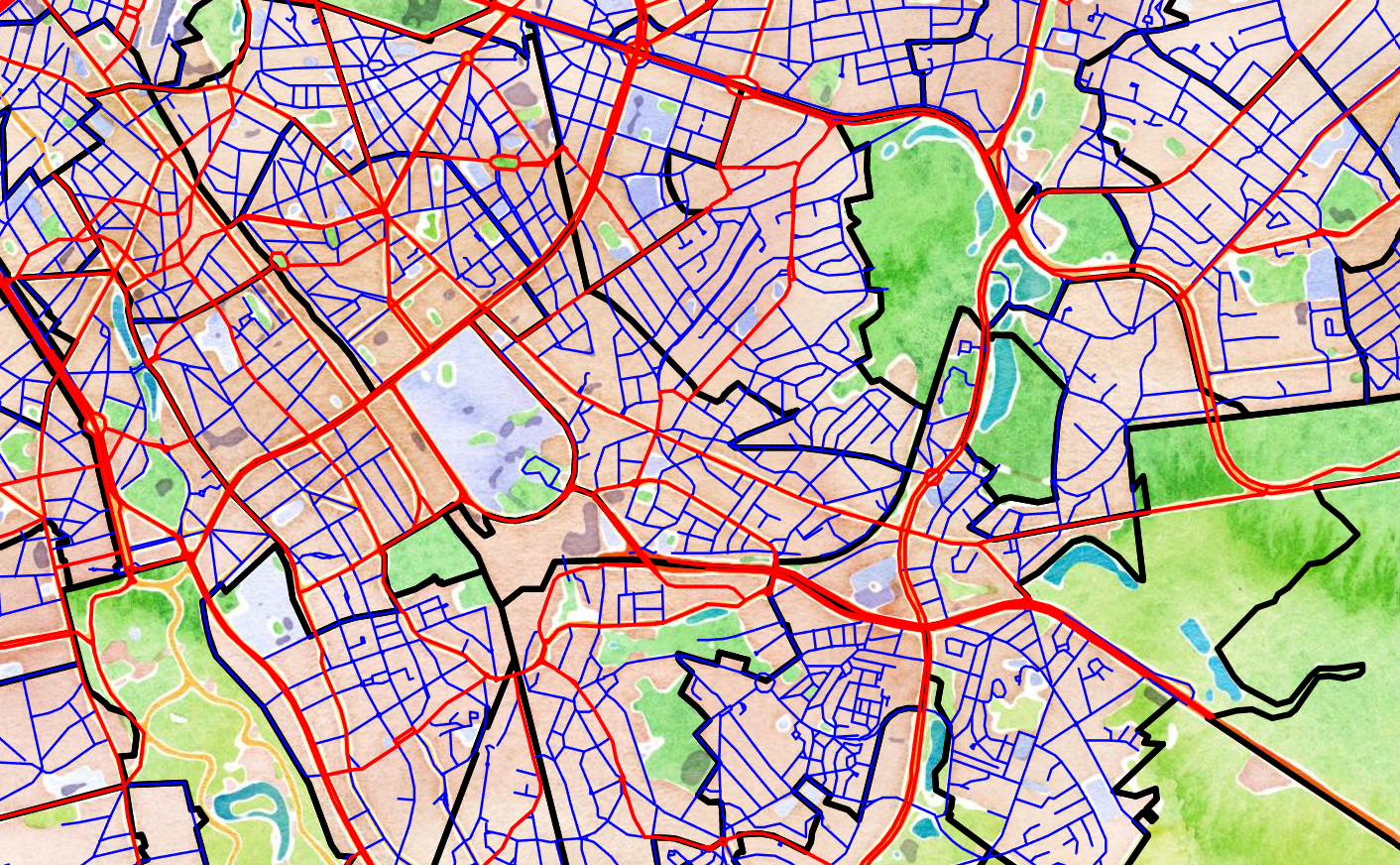}
    \caption{\small Zoom on highways extracted from OSM for Brussels. The blue lines are residential streets, whereas the red ones are major city highways. The bold black lines represent the neighborhoods' boundaries. \break \vspace{-0.3cm} \newline   \centerline{\href{https://rawcdn.githack.com/victor-radermecker/BrusselsEVS35/80081458931974ae7cfe9f19cb79183d21a79034/img/lanes-limitations-brussels.html}{\textsc{Interactive Figure \faGithub}}}}
  \label{fig:lanes-limitations}
\end{wrapfigure}
The first step of the segmentation takes advantage of the city's highways because they are accurately labeled in OSM. As the goal is to derive the proportion of residential charging per neighborhood, we compute the ratio of the total length of residential lanes over the full length of all other roads. 

\vspace{0.5cm}

\autoref{fig:lanes-limitations} shows all the available highways for the whole city. The blue lines represent residential highways, whereas the red ones are motorways, primary, secondary or tertiary highways (major city highways). The bold black lines represent the neighborhood's boundaries. On the one hand, we can identify industrial areas without any residential lanes, for instance, in the city center. On the other hand, we can also recognize some strongly residential neighborhoods. As many roads spread over different districts, a binary search algorithm is used to efficiently split them into sub-lanes fully included in a single neighborhood. 

\subsubsection{Non Residential Segmentation}\label{nonresidential_segmentation}

In the city's large peripheral commercial areas, people often come for a couple of shopping hours and then leave. Therefore, building normal charging stations, which require about 10-12 hours of charging for most vehicles, would not make sense. The most appropriate solution would be to build semi-rapid chargers. A fuel station on a motorway would benefit from fast-charging stations to decrease the drivers' range anxiety. In areas abundant with restaurants, bars, and tourist spots, many people flock for short periods of time. Therefore, a fast-charging station would also probably be the most appropriate solution. In a working district, where drivers park their cars for the whole working day, normal charging points would be the most appropriate. \\

For these reasons, exploring the cities' POIs distribution may provide valuable insights regarding EV charging demand segmentation. We have extracted all POIs available in OSM for Brussels in this work. Then, we have assigned a specific charging technology to each type of POI. \autoref{tab:POI_Table} represents all the points extracted. The investigated amenities range from education (school, universities),  sustenance (bars, restaurants, cafes), shops (convenience stores, supermarkets, shopping malls), to hospitals. \\


\begin{table}[h!]
    \centering
    \captionsetup{width=.96\textwidth}
    \small
    \caption{This table summarizes the different amenities used in the study. The 'Charger' column shows which charging technology was assigned, among normal, semi-rapid, and fast charging. Amenities are grouped into different sectors named in the 'Sector' column. The 'Type of Building' column indicates which POI is discussed. The 'OSM Features' column denotes the specific tag used on OSM to extract the data. The 'Total' column gives the number of POI extracted while the 'Area' indicates the sum of their areas in $km^2$.}
    \label{tab:POI_Table}

\begin{tabular}{l||l|l|l|l|l}
Charger &
  Sector &
  Type of building &
  OSM Features &
  Total &
  Area \\ \hline \hline
 &
  \cellcolor[HTML]{F2F2F2}Home &
  \cellcolor[HTML]{F2F2F2}Residential   areas &
  \cellcolor[HTML]{F2F2F2}highway =   “residential” &
  \cellcolor[HTML]{F2F2F2}NA &
  \cellcolor[HTML]{F2F2F2}NA \\
\multirow{-2}{*}{\begin{tabular}[c]{@{}l@{}}Normal \\    charging\end{tabular}} &
  \cellcolor[HTML]{FFFFFF}Work &
  \cellcolor[HTML]{FFFFFF}Office areas &
  \cellcolor[HTML]{FFFFFF}Amenity =   "office" &
  \cellcolor[HTML]{FFFFFF}1 094 &
  \cellcolor[HTML]{FFFFFF}0.82 \\ \hline
 &
  \cellcolor[HTML]{F2F2F2} &
  \cellcolor[HTML]{F2F2F2}Universities &
  \cellcolor[HTML]{F2F2F2}amenity = “university” &
  \cellcolor[HTML]{F2F2F2} &
  \cellcolor[HTML]{F2F2F2} \\
 &
  \cellcolor[HTML]{F2F2F2} &
  \cellcolor[HTML]{F2F2F2}Schools &
  \cellcolor[HTML]{F2F2F2}amenity =   “school” &
  \cellcolor[HTML]{F2F2F2} &
  \cellcolor[HTML]{F2F2F2} \\
 &
  \multirow{-3}{*}{\cellcolor[HTML]{F2F2F2}Education} &
  \cellcolor[HTML]{F2F2F2}Kindergarten &
  \cellcolor[HTML]{F2F2F2}amenity =   “kindergarten” &
  \multirow{-3}{*}{\cellcolor[HTML]{F2F2F2}606} &
  \multirow{-3}{*}{\cellcolor[HTML]{F2F2F2}5.39} \\
 &
  \cellcolor[HTML]{FFFFFF} &
  \cellcolor[HTML]{FFFFFF}Music Schools &
  \cellcolor[HTML]{FFFFFF} &
  \cellcolor[HTML]{FFFFFF} &
  \cellcolor[HTML]{FFFFFF} \\
 &
  \cellcolor[HTML]{FFFFFF} &
  \cellcolor[HTML]{FFFFFF}Cinemas &
  \cellcolor[HTML]{FFFFFF} &
  \cellcolor[HTML]{FFFFFF} &
  \cellcolor[HTML]{FFFFFF} \\
 &
  \multirow{-3}{*}{\cellcolor[HTML]{FFFFFF}Entertainment} &
  \cellcolor[HTML]{FFFFFF}Theaters &
  \multirow{-3}{*}{\cellcolor[HTML]{FFFFFF}amenity =   “entertainment”} &
  \multirow{-3}{*}{\cellcolor[HTML]{FFFFFF}451} &
  \multirow{-3}{*}{\cellcolor[HTML]{FFFFFF}0.49} \\
 &
  \cellcolor[HTML]{F2F2F2} &
  \cellcolor[HTML]{F2F2F2}Supermarkets &
  \cellcolor[HTML]{F2F2F2} &
  \cellcolor[HTML]{F2F2F2} &
  \cellcolor[HTML]{F2F2F2} \\
 &
  \cellcolor[HTML]{F2F2F2} &
  \cellcolor[HTML]{F2F2F2}Furnitures &
  \cellcolor[HTML]{F2F2F2} &
  \cellcolor[HTML]{F2F2F2} &
  \cellcolor[HTML]{F2F2F2} \\
 &
  \cellcolor[HTML]{F2F2F2} &
  \cellcolor[HTML]{F2F2F2}Health \&   Beauty &
  \cellcolor[HTML]{F2F2F2} &
  \cellcolor[HTML]{F2F2F2} &
  \cellcolor[HTML]{F2F2F2} \\
 &
  \cellcolor[HTML]{F2F2F2} &
  \cellcolor[HTML]{F2F2F2}Clothing &
  \cellcolor[HTML]{F2F2F2} &
  \cellcolor[HTML]{F2F2F2} &
  \cellcolor[HTML]{F2F2F2} \\
 &
  \multirow{-5}{*}{\cellcolor[HTML]{F2F2F2}Shops} &
  \cellcolor[HTML]{F2F2F2}Other &
  \multirow{-5}{*}{\cellcolor[HTML]{F2F2F2}amenity =   "shops"} &
  \multirow{-5}{*}{\cellcolor[HTML]{F2F2F2}4 510} &
  \multirow{-5}{*}{\cellcolor[HTML]{F2F2F2}1.29} \\
 &
  \cellcolor[HTML]{FFFFFF} &
  \cellcolor[HTML]{FFFFFF}Hospital &
  \cellcolor[HTML]{FFFFFF}amenity =   “clinic” &
  \cellcolor[HTML]{FFFFFF}19 &
  \cellcolor[HTML]{FFFFFF} \\
 &
  \multirow{-2}{*}{\cellcolor[HTML]{FFFFFF}Healthcare} &
  \cellcolor[HTML]{FFFFFF}Clinic &
  \cellcolor[HTML]{FFFFFF}amenity =   “hospital” &
  \cellcolor[HTML]{FFFFFF}50 &
  \multirow{-2}{*}{\cellcolor[HTML]{FFFFFF}1.21} \\
\multirow{-14}{*}{\begin{tabular}[c]{@{}l@{}}Semi-Rapid \\    charging\end{tabular}} &
  \cellcolor[HTML]{F2F2F2}Sports &
  \cellcolor[HTML]{F2F2F2}Sport   facilities &
  \cellcolor[HTML]{F2F2F2}amenity =   "sport" &
  \cellcolor[HTML]{F2F2F2}808 &
  \cellcolor[HTML]{F2F2F2}1.24 \\ \hline
 &
  \cellcolor[HTML]{FFFFFF} &
  \cellcolor[HTML]{FFFFFF}Bars &
  \cellcolor[HTML]{FFFFFF}amenity =   “bar” &
  \cellcolor[HTML]{FFFFFF}802 &
  \cellcolor[HTML]{FFFFFF} \\
 &
  \cellcolor[HTML]{FFFFFF} &
  \cellcolor[HTML]{FFFFFF}Cafes &
  \cellcolor[HTML]{FFFFFF}amenity =   “café” &
  \cellcolor[HTML]{FFFFFF}569 &
  \cellcolor[HTML]{FFFFFF} \\
 &
  \cellcolor[HTML]{FFFFFF} &
  \cellcolor[HTML]{FFFFFF}Fast foods &
  \cellcolor[HTML]{FFFFFF}amenity =   “fast\_food” &
  \cellcolor[HTML]{FFFFFF}917 &
  \cellcolor[HTML]{FFFFFF} \\
 &
  \cellcolor[HTML]{FFFFFF} &
  \cellcolor[HTML]{FFFFFF}Ice Cream &
  \cellcolor[HTML]{FFFFFF}amenity =   “ice\_cream” &
  \cellcolor[HTML]{FFFFFF}30 &
  \cellcolor[HTML]{FFFFFF} \\
 &
  \cellcolor[HTML]{FFFFFF} &
  \cellcolor[HTML]{FFFFFF}Pub &
  \cellcolor[HTML]{FFFFFF}amenity =   “pub” &
  \cellcolor[HTML]{FFFFFF}1 136 &
  \cellcolor[HTML]{FFFFFF} \\
 &
  \multirow{-6}{*}{\cellcolor[HTML]{FFFFFF}Sustenance} &
  \cellcolor[HTML]{FFFFFF}Restaurant &
  \cellcolor[HTML]{FFFFFF}amenity =   “restaurant” &
  \cellcolor[HTML]{FFFFFF}3 057 &
  \multirow{-6}{*}{\cellcolor[HTML]{FFFFFF}0.54} \\
 &
  \cellcolor[HTML]{F2F2F2}Tourism &
  \cellcolor[HTML]{F2F2F2}Monuments, Hotels,   etc &
  \cellcolor[HTML]{F2F2F2}amenity =   "tourism" &
  \cellcolor[HTML]{F2F2F2}1 231 &
  \cellcolor[HTML]{F2F2F2}1.31 \\
\multirow{-8}{*}{\begin{tabular}[c]{@{}l@{}}Fast/Ultra-fast \\    charging\end{tabular}} &
  \cellcolor[HTML]{FFFFFF}Taxi &
  \cellcolor[HTML]{FFFFFF}Specific   areas &
  \cellcolor[HTML]{FFFFFF}amenity =   “taxi” &
  \cellcolor[HTML]{FFFFFF}669 &
  \cellcolor[HTML]{FFFFFF}0.007 \\ \hline
\end{tabular}

\end{table}

When all POIs are extracted, a challenge remains. We aim to estimate the number of charging sessions at each POI, and therefore, we need to assess each POI's popularity. 
The optimal solution would be to use the attendance rate of each restaurant, bar, hotel. Such data is only reliably available on one web maps service, GoogleMaps, which is unfortunately quite expensive. This model assumes that the popularity is proportional to the amenity's area. Of course, this is a strong assumption that does not always hold. Sports facilities are great examples of this. A medium-sized supermarket attracts more people daily than a colossal athletics stadium. For this reason, sports facilities were not taken into consideration in the final model, even if they are presented in \autoref{tab:POI_Table}. In practice, different models were explored by researchers in the literature, the latest of which is the \textit{Universal visitation law of human mobility} proposed by Schläpfer, which rigorously proves the aforementioned assumptions \cite{Z10}. If attendance data were to become available, the accuracy of the model could probably be increased considerably. \\

\begin{wrapfigure}[15]{r}{.555\textwidth}
\vspace{-0.25cm}
\begin{align}
  \Phi_{Nres} &= \alpha (1-\gamma) (\Delta_i + \Delta_j) \\
  \Phi_{Noff} &= \frac{a_{office}}{\sum_{i} a_{i}} (1-\alpha)(1-\gamma) (\Delta_i + \Delta_j) \\
  \Phi_{Sem} &=  \frac{a_{semi\_rapid}}{\sum_{i} a_{i}} (1-\alpha) (1-\gamma)  (\Delta_i + \Delta_j) \\
  \Phi_{Rap} &= \frac{a_{rapid}}{\sum_{i} a_{i}}  (1-\alpha) (1-\gamma)  (\Delta_i + \Delta_j) \\
  \Phi_{Par} &= \gamma (\Delta_i + \Delta_j) \\
\end{align}
\end{wrapfigure}
The final charging demand $\Phi$ is computed using the following formulas for each category. $\Phi_{Nres}$ and $\Phi_{Noff}$ are the proportion of normal charging for residential and office purposes respectively. $\Phi_{Sem}$ and $\Phi_{Rap}$ are, respectively, the semi-rapid and fast charging needs. In these equations, $\alpha$ represents the proportion of residential demand  ($\alpha \in [0,1]$), $\gamma$ is the PPR ($\gamma \in [0,1]$), $a_{i}$ is the total area of all POIs assigned to charging type $i$ (Units: $km^2$), $\sum_{i} a_{i}$ is the total area of all POIs within one neighborhood (Units: $km^2$) and finally, $\Delta_i$ and $\Delta_j$ are respectively the regular and irregular charging demand (Units: $kWh$). The DR was taken into consideration when computing $\Delta_i$ and $\Delta_j$.



\section{Results}

\subsection{Case study of Brussels}

\subsubsection{Estimation Results}\label{estimation_results}

\begin{wrapfigure}[33]{r}{0.5\textwidth}
    \includegraphics[width=1.0\linewidth,trim=0pt 0pt 0pt 30pt]{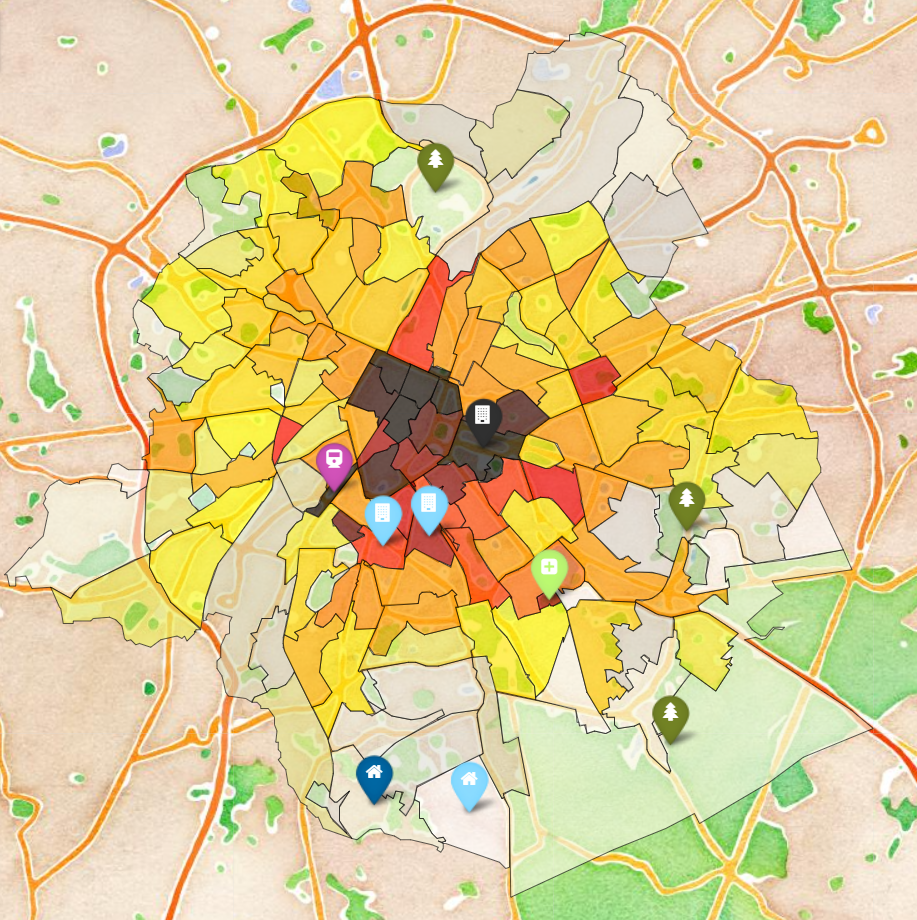}
    \includegraphics[width=1.0\linewidth,trim=0pt 0pt 0pt 0pt]{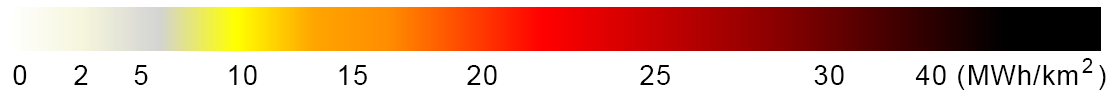}

    \caption{Total Public EV Charging demand normalized by the neighborhood's areas (both regular and irregular trips aggregated). The various markers were added to attract the reader's attention. Addition information about them is given in the core text. Units: $MWh/km^2$.  \break \vspace{-0.2cm} \newline 
    \centerline{ \href{https://rawcdn.githack.com/victor-radermecker/BrusselsEVS35/981a6e68a5b8f22aefaa548aa7c724f556705495/img/Final_TotalEnergyDemandByArea.html}{\textsc{Interactive Figure \faGithub}}}}
    \label{fig:final_demand_by_area}
\end{wrapfigure}

\autoref{fig:final_demand_by_area} shows the final EV public charging demand estimation normalized by each neighborhood's area (in $MWh/km^2$). We can notice that the city center's demand is overall more important than in the suburbs. This is a direct effect of the PPR correction factor. \\

The two blue-home (\textcolor{lightblue}{{\footnotesize \faIcon[regular]{home}}},\textcolor{darkblue}{{\footnotesize \faIcon[regular]{home}}}) neighborhoods in the south are both highly residential. Nevertheless, there is a significant charging demand difference between them, which may, at first, seem paradoxical. The left one, \textit{Homborch}, exceeds 5,2 MWh (around 50 stations needed) whereas the right one, \textit{Viver d'Oie}, has the lowest charging demand of all districts, 0 MWh. How is this difference possible? \textit{Vivier d'Oie} accommodates mainly large detached houses, which leads to a PPR close to $1$ and makes it independent from public infrastructures. On the contrary, \textit{Homborch} accommodates subsidized housings with a low PPR, which justifies the higher density of public chargers. Note that a null charging demand for \textit{Viver d'Oie} is probably unrealistic as under-estimated. The demand is set to 0 because the PPR $\approx 1.03$ means more private parking lots than households. This extreme example highlights the importance of the PPR. If we want the chargers to be ultimately homogeneously dispatched around the city, the PPR value may be capped at 95\% to prevent such situations. 

\vspace{0.5cm}

The black building icon (\textcolor{black}{{\footnotesize \faIcon[regular]{building}}}) in the city center represents the \textit{Quartier Européen}, a working district. It's high charging demand makes sense, but, as we'll show in \autoref{segmentation_results}, companies may partly support it. The two light-blue apartment (\textcolor{lightblue}{{\footnotesize \faIcon[regular]{building}}}) icons show residential areas where many people live in apartments with a very low PPR Ratio. The light-green cross (\textcolor{lightgreen}{{\footnotesize \faIcon[regular]{plus-square}}}) icon represents the Delta neighborhood, hosting the city's largest hospital. The purple-train icon (\textcolor{purple}{{\footnotesize \faIcon[solid]{train}}}) shows a bustling district that includes one of the city's main train stations in Brussels. In such areas, it is understandable that charging needs will be higher. Finally, the green-three icons (\textcolor{darkgreen}{{\footnotesize \faIcon[regular]{tree}}}) symbolize poorly-populated district, such as parks, forests, or cemeteries, which explains the poor (or null) charging demand (white tiles). \\

\subsubsection{Segmentation Results}\label{segmentation_results}

\begin{SCfigure}[50][h!]
    \small
    \caption{
    Segmentation results for three different neighborhoods (\textit{Chant d'Oiseau}, \textit{Matonge}, \textit{Quartier Européen}). The y-axis shows the percentage of each charging technology. 
    \textit{Chant d'Oiseau} is the perfect residential neighborhood. More than 90$\%$ of its charging needs are supported either by normal residential stations or private chargers.  
    \textit{Matonge} is a residential neighborhood with heavy parking restrictions (less than 5$\%$ of private chargers), which also include a substantial number of shops and restaurants (17.3$\%$ of semi-rapid and fast charging needs).
    \textit{Quartier Européen} is a working district, as shown by the 34.4$\%$ of normal office charging. The numerous highways and restaurants justify the 9.5\% need for semi-rapid and fast charging.
    It is interesting to compare these proportions with
    \autoref{fig:segmentation_results}, which highlights the different POIs included in each neighborhood.} 
    \includegraphics[width=0.50\textwidth]{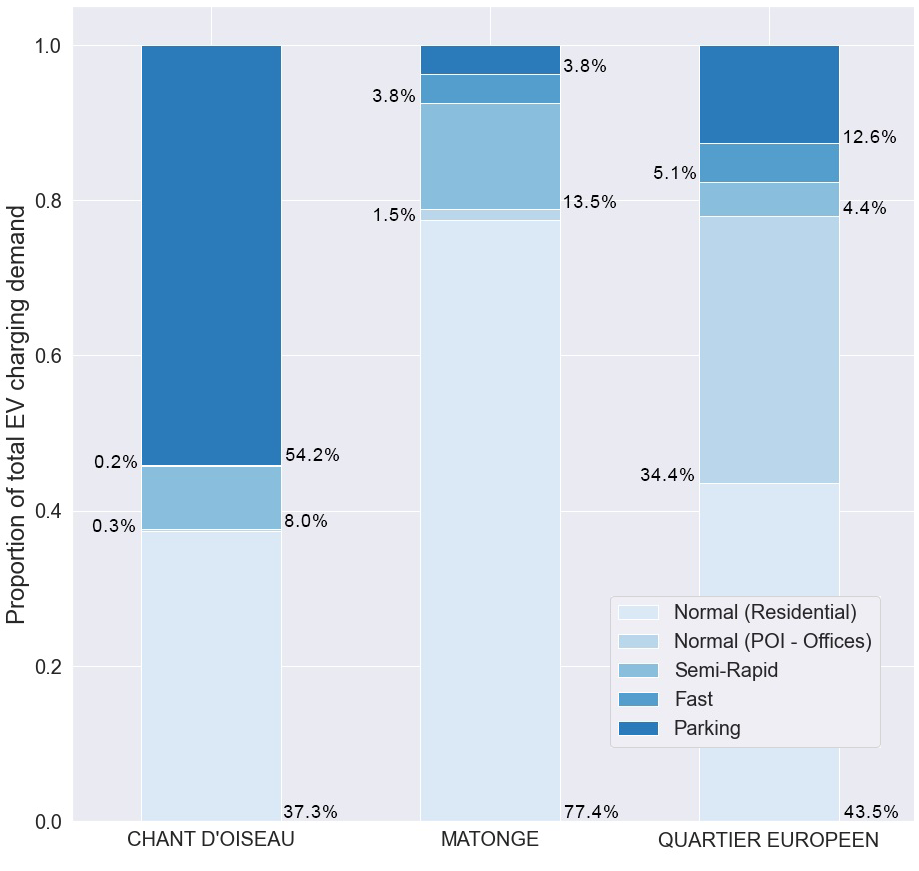}
    \label{fig:segmentations-proportions}
\end{SCfigure}

Let's analyze the segmentation results for three neighborhoods of relatively different nature: \textit{Matonge}, \textit{Chant d'Oiseau}, and \textit{Quartier Européen}. \autoref{fig:segmentations-proportions} shows the final segmentation for these neighborhoods. The total EV charging demand is segmented into Normal (Residential), Normal (POI - Offices), Semi-Rapid, Rapid, and Parking. Parking represents the proportion of demand that drivers' private charging points should support. \\ 

First of all, we can see on \autoref{fig:segmentations-proportions} that most of the charging demand is endorsed by normal and private charging, which means that semi-rapid and rapid charging accounts only for a minority. This is reassuring as these technologies should only be used when the cheaper normal charging solutions are not possible. \\

The normal (POI -  Offices) demand may partly be considered as private charging, as companies will invest in charging points for their employees. European institutions in Brussels have already started massively investing in various EV charging solutions. Therefore, in work districts such as \textit{Quartier Européen}, where the normal (POI - Offices) charging demand is high, the public demand could be reduced. \\

\begin{figure}[!ht] \small
\begin{centering}
  \begin{minipage}[t]{0.315\linewidth}
    \centering
    \includegraphics[width=\linewidth]{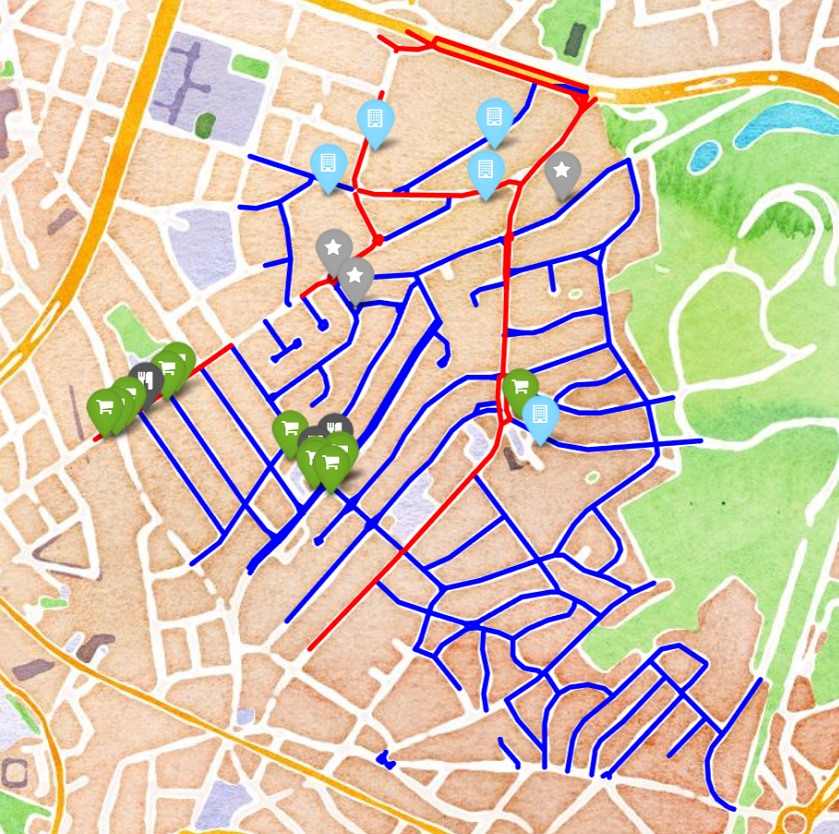} 
    \subcaption{\small The map highlights \textit{Chant d'Oiseau}, a heavily residential neighborhood, as shown by the numerous blue lines, testifying for the high proportion of housing. Regarding public amenities, it only accommodates a few shops. The PPR rate is relatively high due to the low parking regulations.
    \break \vspace{-0.2cm} \newline 
    \centerline{ \href{https://rawcdn.githack.com/victor-radermecker/BrusselsEVS35/de9ea9b64cdebacd508292d16a903faa889c2510/img/neighborhoods-chant-oiseau.html}{\textsc{Interactive Figure \faGithub}}}} 
    \label{ChantD'oiseau}
  \end{minipage}
    \hfill 
    \begin{minipage}[t]{0.315\linewidth}
    \centering
    \includegraphics[width=\linewidth]{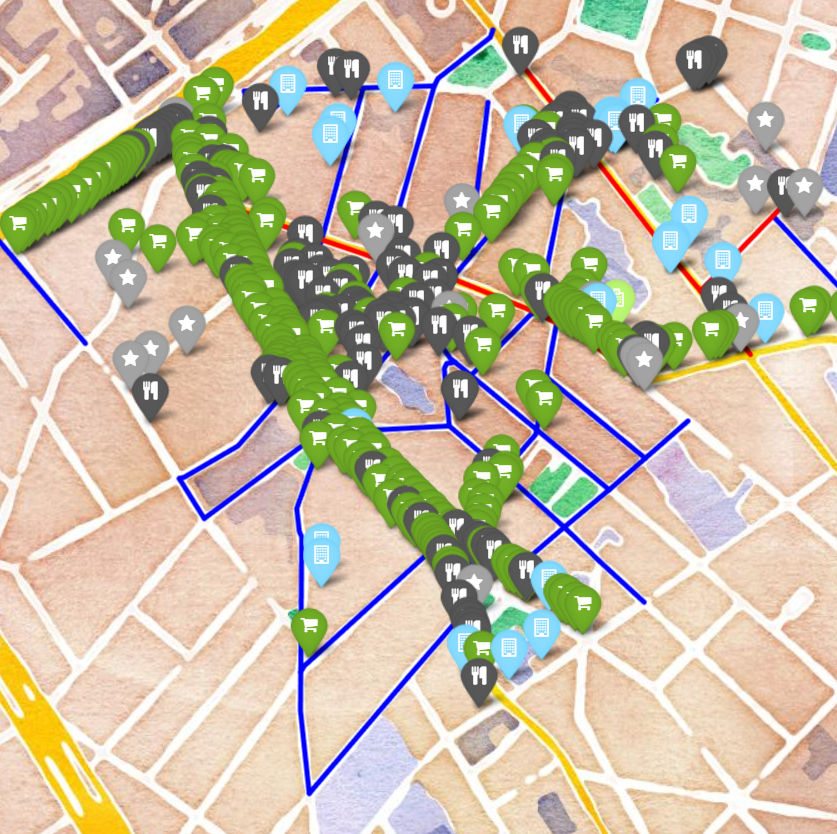}
    \subcaption{\small The map highlights \textit{Matonge}, a neighborhood accommodating many public amenities such as bars, restaurants, and shops (ranging from convenience stores to supermarkets). The high number of blue lines indicates a high proportion of housing. The meager PPR rate is due to heavy parking restrictions.
    \break \vspace{-0.2cm} \newline 
    \centerline{ \href{https://rawcdn.githack.com/victor-radermecker/BrusselsEVS35/88042091b57ba1b6f3d2e1e3765e7dec4ebecb47/img/neighborhoods-matonge.html}{\textsc{Interactive Figure \faGithub}}} }
    \label{Matonge}
  \end{minipage} 
  \hfill 
  \begin{minipage}[t]{0.315\linewidth}
    \centering
    \includegraphics[width=\linewidth]{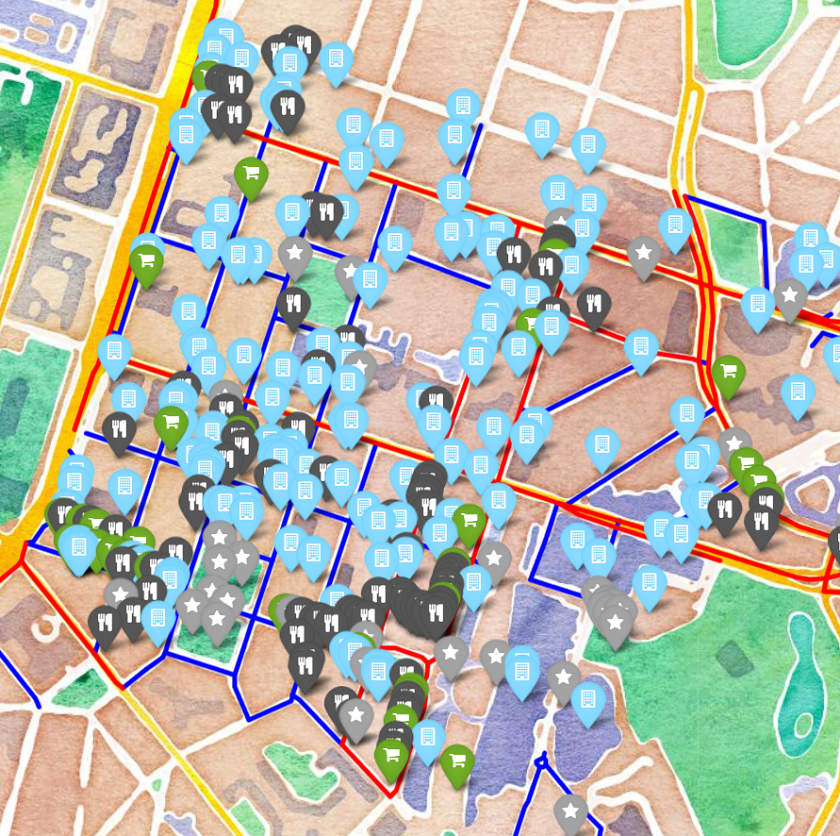} 
    \subcaption{\small The map highlights \textit{Quartier Européen}, a working district which accommodates many offices (mainly the buildings of the European Institutions). It abounds with restaurants and includes many long-term parking lots as people park their cars for the whole day while working. Some streets also have housing.
    \break \vspace{-0.2cm} \newline 
    \centerline{
    \href{https://rawcdn.githack.com/victor-radermecker/BrusselsEVS35/de9ea9b64cdebacd508292d16a903faa889c2510/img/neighborhoods-quartier-europ\%C3\%A9en.html}{\textsc{Interactive Figure \faGithub}}}}
    \label{Europe}
  \end{minipage}

  \centerline{
    \begin{minipage}[b]{0.75\textwidth}
    \centerline{\includegraphics[width=0.9\linewidth]{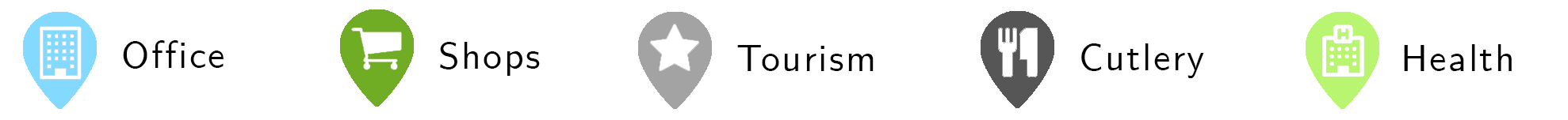}}
    \end{minipage}}
  \centerline{
    \begin{minipage}[b]{0.75\textwidth}
    \centerline{\includegraphics[width=0.9\linewidth]{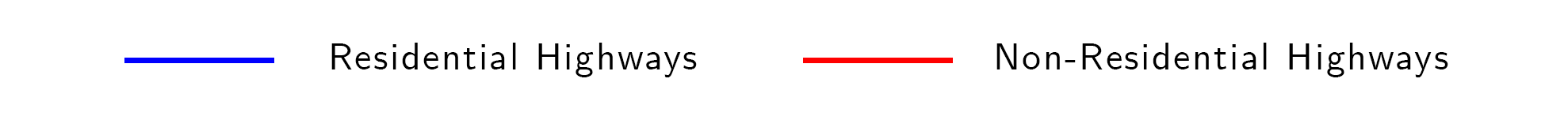}}
    \end{minipage}}
    
    \caption{\small This figure shows three neighborhoods, each of them having a different nature. Their segmentation results are presented in \autoref{fig:segmentations-proportions}. All markers indicate a specific amenity extracted in OSM.}
    \label{fig:segmentation_results}
\end{centering}
\end{figure}

\autoref{fig:segmentation_results} shows the different amenities accessible within each of these neighborhoods. We can distinguish the different nature of each neighborhood. Many blue lines indicate a high residential proportion. Areas including many public facilities, such as restaurants, supermarkets and tourist spots, are interesting places for semi-rapid or rapid charging stations.

\newpage

\subsubsection{City-level Results}

\begin{wraptable}[19]{r}{7cm}
    \small
    \vspace{-0.7cm}
	\caption{Assumptions taken to convert the demand in $kWh$ into a number of stations \cite{A2}.}
	\centering
	\begin{tabular}{l||l|l}
        Charger & 
        Assumption&
        Value \\
        \hline
        \hline 
	
		\multirow{2}*{Normal} & Theoretical Power & 7 kW \\
		\cline{2-3}
		~ & Avg. Pow. Delivery & 80\% \\
		\cline{2-3}
		~ & Occupancy Rate & 50\% \\
		\cline{2-3}
		~ & Opening hours & 24h/24h \\
		\hline

		\multirow{2}*{Semi-Rapid} & Theoretical Power & 22 kW \\
		\cline{2-3}
		~ & Avg. Pow. Delivery & 80\% \\
		\cline{2-3}
		~ & Occupancy Rate & 80\% \\
		\cline{2-3}
		~ & Opening hours & 10h-20h \\
		\hline
		
		\multirow{2}*{Rapid} & Theoretical Power & 100 kW \\
		\cline{2-3}
		~ & Avg. Pow. Delivery & 80\% \\
		\cline{2-3}
		~ & Occupancy Rate & 25 \% \\
		\cline{2-3}
		~ & Opening hours & 24h/24h \\
	
        \hline
		\hline
	\end{tabular}
	\label{tab:assumptions}
\end{wraptable}
Having estimated and segmented the demand into different charging technologies, it is interesting to convert these energy values ($kWh$) into a number of charging stations. \autoref{tab:assumptions} summarizes the different assumptions made for this conversion. Most of them are based on an analysis proposed by Brussels Environment, the city planning authorities in Brussels \cite{A2}. 

\vspace{0.4cm}

For each technology, we consider a value for the theoretical power (maximal power deliverable by the terminal), average power delivery (mean power delivered over the whole charging session), occupancy rate (proportion of time when a vehicle is plugged-in and charging), and opening hours (the time when the station is open for charging).

\vspace{0.4cm}

\autoref{tab:nbr-charging-stations} gives the final number of charging stations needed to cover the whole city. We get 11 766 normal charging stations for residential purposes only. Regarding at-office charging stations, the 1 047 stations may be, at least partly, supported by companies. This would substantially reduce the pressure on public services. Around 3 004 semi-rapid charging stations are needed to efficiently cover the demand around POIs such as universities, museums, and commercial centers. Finally, let's consider that only normal charging stations should meet the needs of the entire charging demand, which is the currently preferred approach of Brussels Environment. We then get 24 413 stations for the whole city. If we add the Good Move ambitions, a plan aiming to reduce traffic in the city \cite{A2} by at least 5\%, this gives 23 193 points. \\

\begin{wraptable}[11]{r}{5cm}
\small
\vspace{-0.75cm}
    \caption{Number of charging points needed to cover the total demand.}
    \centering
    \begin{tabular}{l||l}
        Charging Type &  Nbr. Stations \\
        \hline
        \hline 
        Normal (Resi) & 11 766 \\
        \hline 
        Normal (Work) & 1 047 \\
        \hline 
        Semi-Rapid & 3 004 \\
        \hline
       Rapid & 312 \\
        \hline
        Full Normal & 24 413 \\
        \hline

    \end{tabular}
    \label{tab:nbr-charging-stations}
\end{wraptable}
This last value of 23 193 charging points aligns with the 22 000 charging points derived by Brussels authorities \cite{A2}.  Results show that focusing solely on one technology may not be optimal, as the number of normal stations is snowballing. Many cities plan to free as much public space as possible. As shown in \autoref{tab:nbr-charging-stations}, mixing a variety of charging technologies can substantially reduce the total number of charging points (16 129 instead of 24 413). Furthermore, semi-rapid and fast charging stations can be efficiently implemented in semi-private areas, such as supermarkets and shopping malls.

\vspace{0.4cm}

\begin{figure}
\centering
\small
    \begin{subfigure}{0.475\linewidth}
    \includegraphics[width=\linewidth]{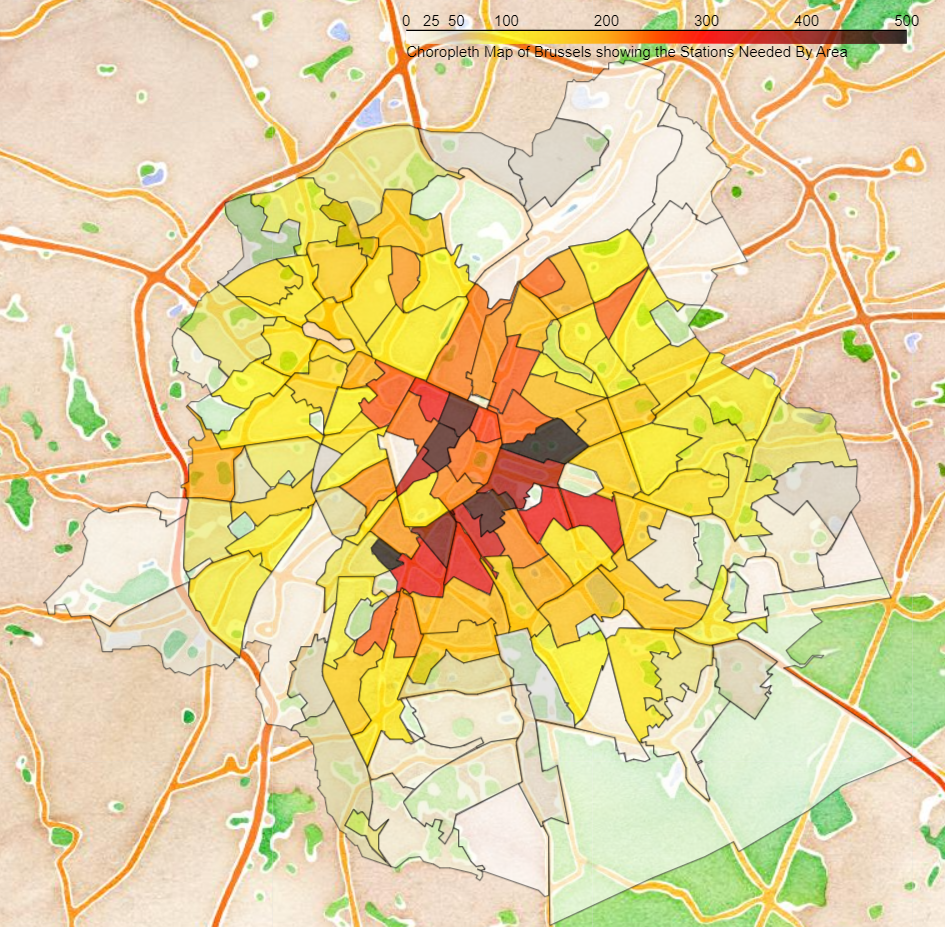}
    \caption{Number of normal (residential) charging stations needed in Brussels normalized by each neighborhood's area.  \href{https://rawcdn.githack.com/victor-radermecker/BrusselsEVS35/ef5e686dffa3040781fadd91c96eefe8cd3b3029/img/NormalStationsNeededByArea.html}{\textsc{Interactive Figure \faGithub}}}
    \label{semi-b}
    \end{subfigure}
    \hfil
    \begin{subfigure}{0.475\linewidth}
        \includegraphics[width=\linewidth]{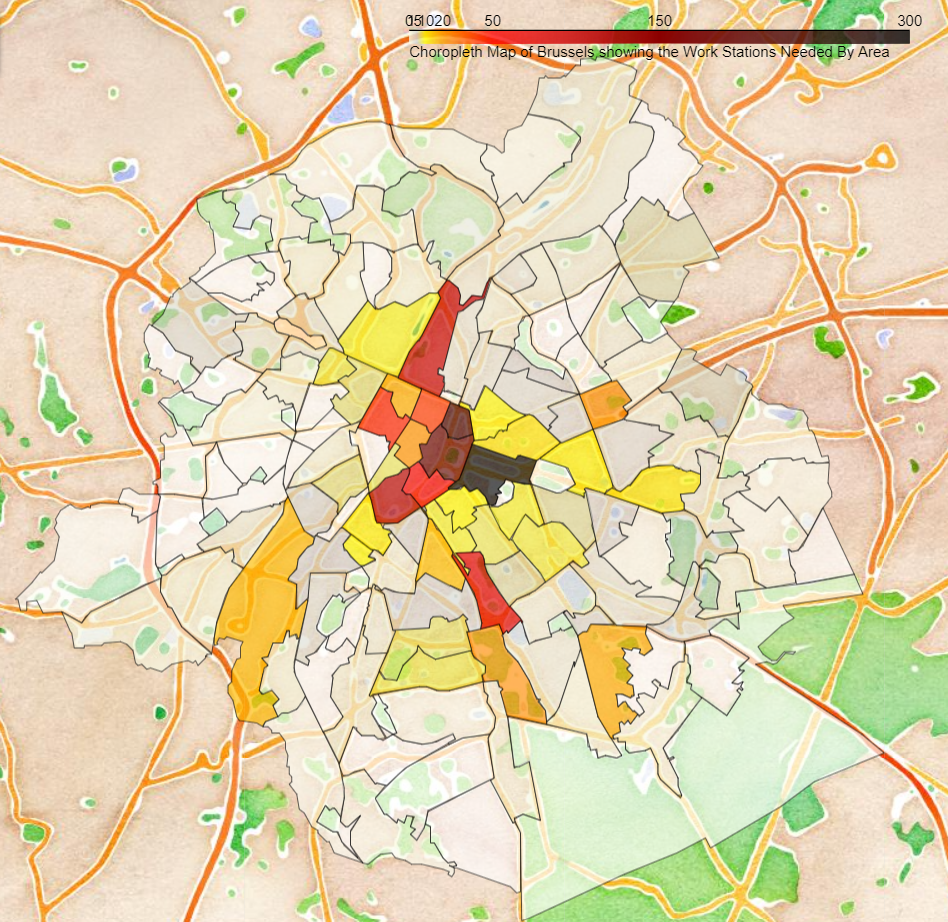}
    \caption{Number of normal at-office charging stations needed in Brussels normalized by each neighborhood's area.  \href{https://rawcdn.githack.com/victor-radermecker/BrusselsEVS35/ef5e686dffa3040781fadd91c96eefe8cd3b3029/img/WorkStationsNeededByArea.html}{\textsc{Interactive Figure \faGithub}}}
    \label{rapid-b}
    \end{subfigure}
    
    \begin{subfigure}{0.475\linewidth}
        \includegraphics[width=\linewidth]{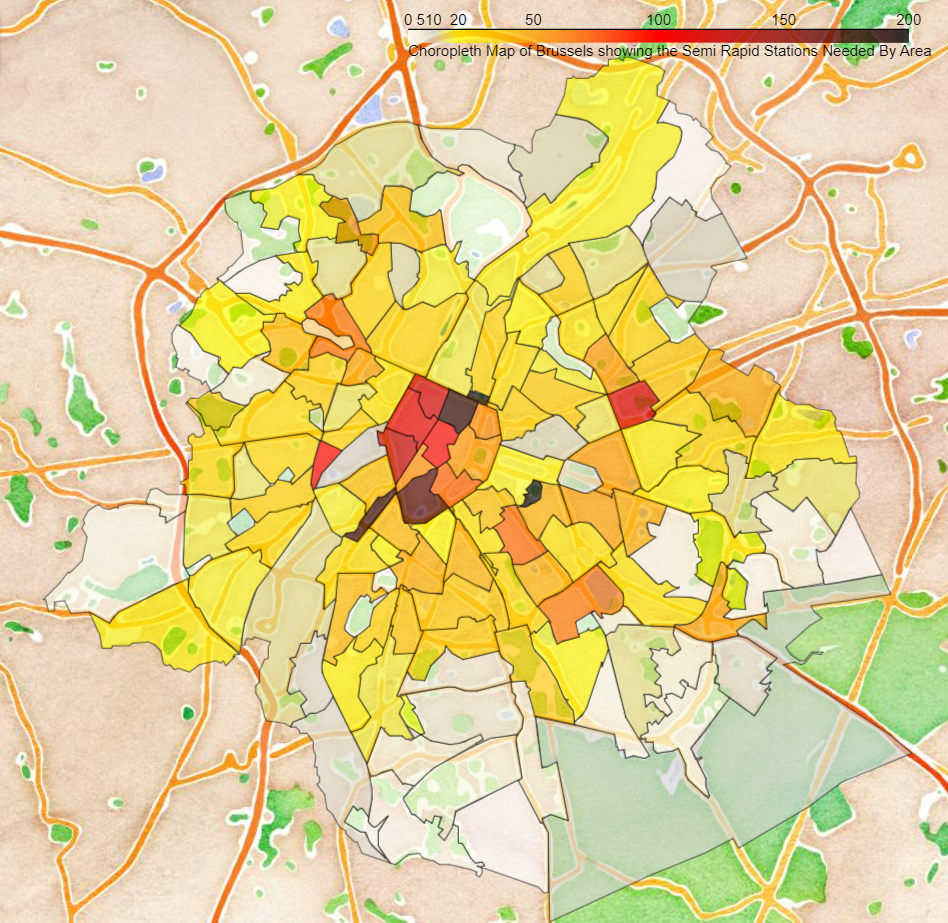}
    \caption{Number of semi-rapid charging stations needed in Brussels normalized by each neighborhood's area. \newline  \href{https://rawcdn.githack.com/victor-radermecker/BrusselsEVS35/ef5e686dffa3040781fadd91c96eefe8cd3b3029/img/Semi\%20RapidStationsNeededByArea.html}{\textsc{Interactive Figure \faGithub}}}
        \label{semi-b}
    \end{subfigure}
    \hfil
    \begin{subfigure}{0.475\linewidth}
        \includegraphics[width=\linewidth]{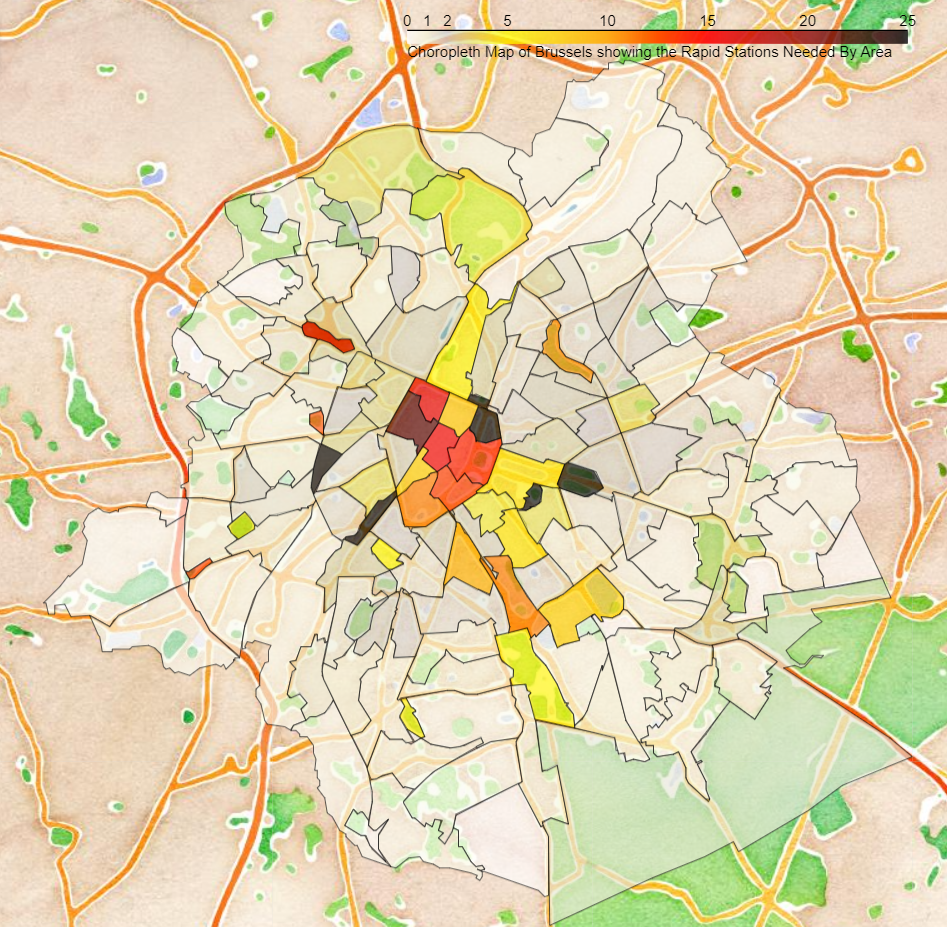}
    \caption{Number of fast charging stations needed in Brussels normalized by each neighborhood's area. \newline \href{https://rawcdn.githack.com/victor-radermecker/BrusselsEVS35/ef5e686dffa3040781fadd91c96eefe8cd3b3029/img/RapidStationsNeededByArea.html}{\textsc{Interactive Figure \faGithub}}}
    \label{rapid-b}
    \end{subfigure}
   
\caption{Number of charging stations needed for each technology. The number of stations are normalized by the neighborhood's area to facilitate data interpretation and visualization. (Units: number of stations / $km^2$).}
    \label{fig:charging}
    \end{figure}

The choropleth maps shown on \autoref{fig:charging} presents the final number of stations needed normalized by the area at neighborhood-level. Some interesting patterns appear. For residential charging, the stations are fairly evenly distributed over the territory. Some sites, even though they are highly residential such as the south of the city, are less dependent on public infrastructure because of the high rate of ownership of private chargers. For office charging, the results seem relevant as we can identify the different city's work districts, such as \textit{Industrie Sud} or \textit{Quartier Européen}. \\

The city's central railway station, \textit{Bruxelles Midi}, shows a high dependence on rapid and semi-rapid stations. It is a mobility hub in Brussels for individual vehicles, shared cars, and taxis, which justify these costly charging technologies. Fast charging seems primarily concentrated in a few downtown areas, which might not be optimal. One ambition of Brussels is to reduce the traffic burden in the city center, and placing fast chargers there would probably increase traffic. Therefore, one should always remain critical of the raw model's results and see if they can be realistically implemented in the city. The best solution would probably be to build these stations along major highways, such as the motorways that surround the city center, called the small belt. \\

Note that the 50\% occupancy assumption is very optimistic, but was used as it is the value assumed by Brussels government \cite{A2}. Different researchers have explored these questions, and it seems that the actual occupancy rates of EV charging stations are much lower. Wolbertus and van den Hoed \cite{Z9} highlighted relatively low occupancy rates for normal charging stations, such as 15\% during the daytime and 25\% during the nighttime. Gnann et al. \cite{Z11} also explored the occupancy rates, for fast charging this time, in Germany and Sweden. Results show that the occupancy rates seem to increase with power delivery, and for $100kW$, rates ranged from 18\% to 28\%.

\newpage

\subsection{Underlying Assumptions and Possible Improvements}

Regarding the demand estimation, three significant assumptions need to be stated. First, this study considers that the proportions of car, walking, and public transport trips are geographically constant and solely depend on the average distance between two neighborhoods. In reality, these proportions will highly depend on the city area. In downtown, public transports are a preferred solution as they are incredibly efficient. In the suburbs, people rely on their private cars as public transport is scarce. Obtaining an OD matrix for public transportation trips in the city would significantly increase the model's accuracy. \\

Second, as this work aggregates the total demand at a neighborhood level, it presumes that people will charge their car after each sufficiently long break. In practice, drivers' charging behavior remains unknown due to immature battery technologies, missing data, and scarce charging infrastructure. Predicting EV drivers' future behavior is an active research topic \cite{charging-behavior}. However, as the goal is here to produce prevailing neighborhood-level trends, such assumptions do not considerably harm the results. \\

Finally, the present work assumes each trip of the input dataset as an EV trip, an assumption that does not hold today as only around 8\% of Brussels cars are electric \cite{A1}. This distribution is assumed constant over the coming years despite Brussels' ambitions to reduce the city's daily traffic.  \\

Regarding the segmentation process, here are the two principal hypotheses. First, the Highway Analysis elaborated in \autoref{residential_segmentation} considers the residential lanes extracted from OSM as fully residential. In reality, this is not always the case. \autoref{Matonge} shows some residential streets which are full of shops and restaurants. Physically, the ground floor is rented for commercial purposes, while the upper floors are housings. Therefore, this analysis probably tends to over-estimate the residential proportion. One solution could be to estimate the ratio of the number of houses over the number of buildings (houses and POIs). \\

Second, the popularity of each POI is assumed proportional to its area. This assumption may not always hold true. Therefore, using the real frequentation rate of each amenity would be the most trustworthy solution.

\section{Conclusion}

The major breakthrough in this article is the EV charging demand segmentation process, which is an approach that researchers have not yet explored. These segmentation results are vital to elaborate roll-out strategies for EV charging infrastructure. By adding the charging speed layer of granularity to the demand, we can apply the optimization models listed in \autoref{independent} to find the optimal locations of each type of charger. Therefore, the model provides actionable insights for city planning authorities and opens the door to more complex deployment road maps. This model is a preliminary version, reasonably straightforward, and supported by rather strong premises. Many improvement directions are possible and would eventually allow obtaining optimal results. \\

The case study proposed proved to be quite precise and one of the model's main advantages is its easy scalability to other cities. Indeed, all prominent cities around the world are rigorously annotated in OSM. Furthermore, our results for Brussels converge to previous studies on the conclusion that mixing various charging technologies is always preferable to only one type \cite{Z6} \cite{Z7} \cite{Z8}. Although associated deployment costs will be higher, it allows to substantially reduce the overall number of charging points and maximize the freed-up public space. Moreover, private companies may partly support these costs, as semi-rapid and fast chargers are meant to be built in semi-private areas, such as supermarkets, shopping malls, etc.

\section*{Acknowledgments}

At the very outset, I would like to thank my supervisor, Professor Lieselot Vanhaverbeke, whose expertise was invaluable in formulating the research questions and methodology. She allowed me to work on a fascinating topic and access datasets that proved critical to the analysis. The door to her office was always open to help me whenever I ran into an issue. The insightful feedback pushed me to sharpen my thinking and brought my work to a higher level. \\

I would also like to express my sincere thanks to all doctoral students of the Electric Vehicle team of the VUB MOBI lab, and especially Simon Weekx, without whom I would not have been able to complete this research. Their numerous advice proofreadings throughout the year were of precious help.

\newpage



\bibliography{content/bibliography.bib}
\bibliographystyle{evs.bst}   

\small

\bigskip
\section*{Presenter Biography}
\begin{minipage}[b]{21mm}
\includegraphics[width=20mm]{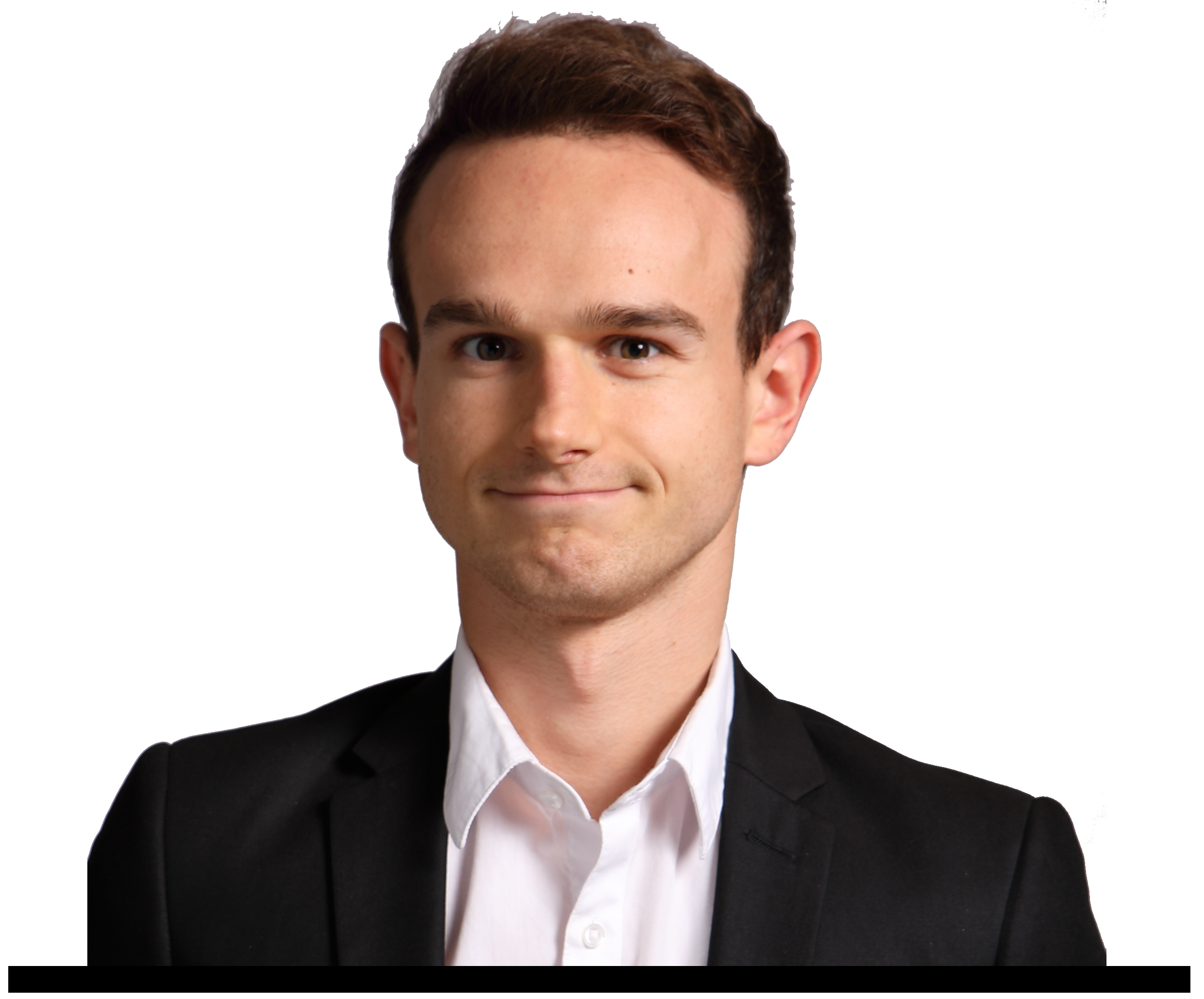}
\end{minipage}
\hfill
\begin{minipage}[b]{140mm}
\small
Victor Radermecker is a Data Science student working on a master thesis in collaboration with Vrije Universiteit Brussel (VUB) and Université Libre de Bruxelles (ULB). Currently graduating from a double degree between ULB and Ecole polytechnique (Paris), his interests range from artificial intelligence to any topic related to renewable energies and sustainable mobility. Next year, he will pursue his studies with a complementary Master of Science at the Massachusetts Institute of Technology (MIT).
\end{minipage}
\\\\

\end{document}